\documentclass[a4paper,12pt]{article}
\usepackage[left=25mm, right=25mm, bottom=35mm, top=35mm]{geometry}

\usepackage[T2A]{fontenc}
\usepackage[utf8]{inputenc}
\usepackage[english]{babel}
\usepackage[pdftex]{graphicx}
\usepackage{amsmath,amssymb,ntheorem}
\usepackage{multicol}
\usepackage{colortbl} %%-added for color tables

\usepackage{geometry}
\usepackage{epsf}
\usepackage{amsfonts}
\usepackage{amsmath}
\usepackage{amssymb}
\usepackage{latexsym}
\usepackage{units}
\usepackage{textcomp}
\usepackage{wrapfig}
\usepackage{booktabs}
\usepackage{hyperref}
\usepackage{xcolor}
\hypersetup{
    colorlinks,
    linkcolor={red!50!black},
    citecolor={blue!50!black},
    urlcolor={blue!80!black}
}

\everymath{\displaystyle}
\usepackage{color}
\everymath{\displaystyle}

\renewcommand{\phi}{\varphi}

\usepackage{url}

\begin{document}
%\begin{titlepage}%-----------------------------------------------------------------------------
%\vspace{-12mm}
%\prepnum{2018-xx}{}
%{
%\renewcommand{\thefootnote}{\textit{\alph{footnote}}}
%\vspace{-25mm}
%\author{\small M.M.~Shapkin, S.A.~Akimenko, A.V.~Artamonov, 
%A.M.~Blik, V.S.~Burtovoy, S.V.~Donskov, A.P.~Filin, A.V.~Inyakin,
%A.M.~Gorin, G.V.~Khaustov, S.A.~Kholodenko,
%V.N.~Kolosov, A.S.~Konstantinov, V.F.~Kurshetsov, V.A.~Lishin,  M.V.~Medynsky,
%Yu.V.~Mikhailov, V.F.~Obraztsov, V.A.~Polyakov, A.V.~Popov, V.I.~Romanovsky,
%V.I.~Rykalin,  A.S.~Sadovsky, V.D.~Samoilenko,  V.K.~Semenov,
%O.V.~Stenyakin,  O.G.~Tchikilev, V.A.~Uvarov, O.P.~Yushchenko\\ 
%\textsc{(NRC "Kurchatov Institute"${}_{}^{}$-${}_{}^{}{}_{}^{}{}_{}^{}$IHEP, Protvino),} \\
%\vspace{1mm}
%V.A.~Duk\instref{a},
%S.N.~Filippov, E.N.~Gushchin, 
%A.A.~Khudyakov, V.I.~Kravtsov,\\ 
%Yu.G.~Kudenko\instref{b}, 
%A.Yu.~Polyarush\\ 
%\textsc{(INR-RAS, Moscow),} \\
%\vspace{1mm}
%V.N.~Bychkov, G.D.~Kekelidze, V.M.~Lysan, B.Zh.~Zalikhanov\\ 
%\textsc{(JINR, Dubna)}}
%\vspace{-2mm}
%\title{\boldmath Study of the decay $K^{+}\to\pi^{+}\pi^{-}\pi^{+}\gamma$ in the OKA experiment}
%\vspace{-4mm}
%\submitted{EPJC}
%\vspace{-14mm}
%\instfoot{a}{\scriptsize{Also~at~University~of~Birmingham,~Birmingham,~United~Kingdom}}
%\instfoot{b}{\scriptsize{Also at Moscow Institute of Physics and Technology, Moscow Region, Russia,\\ and at NRNU Moscow Engineering Physics Institute (MEPhI), Moscow, Russia}}
%}
%\end{titlepage}%-------------------------------------------------------------------------------
~
\vspace{-30mm}
 %\begin{flushright}
 %\footnotesize
 %{\rmfamily\textcolor{black!60}{To be submitted to EPJC}}
 %\end{flushright}
\begin{flushleft}%%{flushright}
\footnotesize  %\scriptsize 
 %\rmfamily      %\it 
 %\ttfamily
 %\textcolor{black!70}{This is a pre-print of an article published in Eur.\hspace{0.5mm}Phys.\hspace{0.5mm}J.\hspace{0.5mm}C\hspace{0.4mm}(2018)\hspace{0.4mm}78:\hspace{0.2mm}92.\\}
 %\textcolor{black!70}{The final authenticated version is available online at: \href{https://doi.org/10.1140/epjc/s10052-018-5566-x}{https://doi.org/10.1140/epjc/s10052-018-5566-x}}\\
 %{\color{gray}\rule{\linewidth}{0.1mm}}
\begin{tabular}[c]{l}
\rowcolor[gray]{.9}This is a pre-print of an article published in Eur.\hspace{0.5mm}Phys.\hspace{0.5mm}J.\hspace{0.5mm}C\hspace{0.4mm}(2019)\hspace{0.4mm}79:\hspace{0.2mm}296.\\
\rowcolor[gray]{.9}The final authenticated version is available online at: \href{https://doi.org/10.1140/epjc/s10052-019-6797-1}{https://doi.org/10.1140/epjc/s10052-019-6797-1}
\end{tabular}
\end{flushleft}%%{flushright}
\vspace{5mm} %\vspace{10mm}
~
\vspace{0mm}
\begin{flushright}
\end{flushright}
\begin{center}
\section*{ \boldmath Study of the decay $K^{+}\to\pi^{+}\pi^{-}\pi^{+}\gamma$ in the OKA experiment}
\end{center}

\begin{center}
{\large 
\textsc{The OKA collaboration}\\
}
\vspace{3mm}

\begin{minipage}{0.9\linewidth}
 \center{
  %\small 
  %\ttfamily
  %\rmfamily
  %\textsf
  \textsc
  M.M.~Shapkin, S.A.~Akimenko, A.V.~Artamonov, 
  A.M.~Blik, V.S.~Burtovoy, S.V.~Donskov, A.P.~Filin, A.V.~Inyakin,
  A.M.~Gorin, G.V.~Khaustov, S.A.~Kholodenko,
  V.N.~Kolosov, A.S.~Konstantinov, V.F.~Kurshetsov, V.A.~Lishin,  M.V.~Medynsky,
  Yu.V.~Mikhailov, V.F.~Obraztsov, V.A.~Polyakov, A.V.~Popov, V.I.~Romanovsky,
  V.I.~Rykalin,  A.S.~Sadovsky, V.D.~Samoilenko,  V.K.~Semenov,
  O.V.~Stenyakin,  O.G.~Tchikilev, V.A.~Uvarov, O.P.~Yushchenko
 }\vspace{-4mm}
 \center{\small 
   %\itshape
   \textsc{(NRC "Kurchatov Institute"${}^{}_{}{}^{}$-${}^{}_{}{}^{}$IHEP, Protvino, Russia),} 
 }\\
 \vspace{-3mm}
 \center{
  %\small
  \rmfamily
  V.A.~Duk\footnote{\scriptsize Now~at~University~of~Birmingham,~Birmingham,~United~Kingdom}, 
  S.N.~Filippov, E.N.~Guschin, 
  A.A.~Khudyakov, V.I.~Kravtsov, 
  %Yu.G.~Kudenko${}^{*}$, 
  %Yu.G.~Kudenko\footnote{\scriptsize Also~at~NRNU~MEPhI,~Moscow,~Russia}, 
  Yu.G.~Kudenko\footnote{\scriptsize Also at Moscow Institute of Physics and Technology, Moscow, Russia}\footnote{\scriptsize Also at NRNU Moscow Engineering Physics Institute (MEPhI), Moscow, Russia}, 
  A.Yu.~Polyarush
 }\vspace{-4mm}
 \center{\small 
   %\itshape 
    \textsc{(INR RAS, Moscow, Russia),}
 }\\
 \vspace{-3mm}
 \center{
  %\small
  \rmfamily
  V.N.~Bychkov, G.D.~Kekelidze, V.M.~Lysan, B.Zh.~Zalikhanov
 }\vspace{-4mm}
 \center{\small 
   \itshape 
   \textsc{(JINR, Dubna, Russia)}\\
 }
\end{minipage}
\end{center}

\vspace{0mm}
\begin{center}
\begin{minipage}{0.09\linewidth}
~
\end{minipage} 
\begin{minipage}{0.78\linewidth}
{ %\sffamily
  \rmfamily
{\bf Abstract.}
  A high statistics data sample of the decays of $K^+$ mesons to three charged particles was accumulated by the OKA experiment in 2012 and 2013. 
  This allowed to select a clean sample of about 450 events with $K^{+}\to\pi^{+}\pi^{-}\pi^{+}\gamma$ decays with the energy of the photon 
  in the kaon rest frame greater than 30 MeV. The measured branching fraction of the $K^{+}\to\pi^{+}\pi^{-}\pi^{+}\gamma$, 
  with $E_{\gamma}^{*}$ > 30 MeV is $(0.71 \pm 0.05) \times 10^{-5}$. The measured energy spectrum of the decay photon is compared 
  with the prediction of the chiral perturbation theory to {\cal O}$(p^{4})$. A search for an up-down asymmetry of the photon 
  with respect to the  hadronic system  decay plane  is also performed.
}
\end{minipage}
\begin{minipage}{0.09\linewidth}
~
\end{minipage}
\end{center}
\vspace{0mm}

\selectlanguage{english}
\setcounter{footnote}{0}

\section{Introduction}\label{SectInitro}
The present experimental status of $K^{+}\to\pi^{+}\pi^{-}\pi^{+}\gamma$ is rather meagre.
It was observed in one experiment on statistics of 7 events \cite{Barmin}. The photon energies in these
events were low, that did not allow to search for deviations from а simple QED process of photon emission. The measured value  of the  $K^{+}\to\pi^{+}\pi^{-}\pi^{+}\gamma$ branching ratio, with 
$E_{\gamma}^{*} > 5 MeV$ is $(1.04\pm0.31) \times 10^{-4}$ \cite{PDG}. In the  present analysis we have
a possibility for more detailed study of this decay using  larger data sample collected by the OKA experiment. The appropriate theory framework for such investigation is the chiral perturbation theory \cite{CHPT} (CHPT).
To the lowest order in an expansion in momenta and meson masses, the radiative decays are completely determined \cite{Ambr1} by the non-radiative amplitude for $K^{+}\to\pi^{+}\pi^{-}\pi^{+}$. At next-to-leading order, a full-fledged CHPT calculation of nonleptonic weak amplitudes of {\cal O}$(p^{4})$ is required. Such a calculation is done in \cite{Ambr2}.
  
\section{Separated kaon beam and OKA experiment}\label{SectExpOverview}
\begin{figure}[!ht]
%\begin{center}
\includegraphics[width=1.0\textwidth]{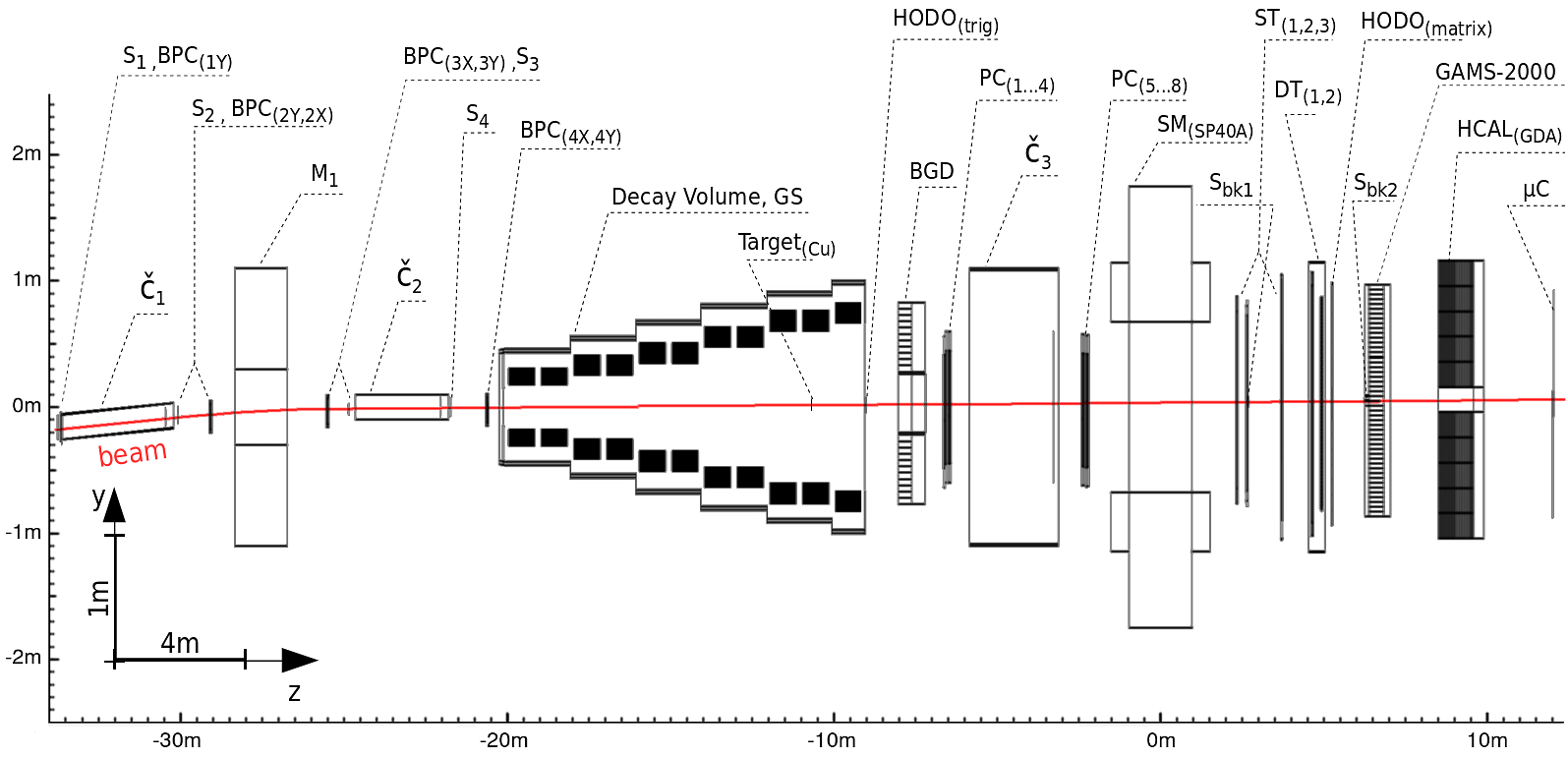}
\caption[OKA setup.]{
    %%\descrfnt
    Schematic elevation view of the OKA setup, see the text for details.
}
\label{FigOkaSetup}
%\end{center}
\end{figure}

The OKA experiment 
makes use of a secondary hadron beam at the U-70 Proton Synchrotron of 
%--IHEP, Protvino, 
NRC "Kurchatov Institute"-$^{}$IHEP, Protvino,
with enhanced fraction of kaons obtained by RF-separation 
with Panofsky scheme \cite{OkaSecondaryBeam}. 
The OKA setup, Fig.~\ref{FigOkaSetup}, is a double magnetic spectrometer complemented by electromagnetic
and hadron calorimeters and a Decay Volume. The first magnetic spectrometer, consisting of the magnet {\small M$_{1}$} and surrounding 1~mm pitch PC's 
({\small BPC$_{(1Y)}$, BPC$_{(2Y,2X)}$}, {\small BPC$_{(3X,3Y)}$}, {\small BPC$_{(4X,4Y)}$}) serves for the  beam momentum measurement.
It is supplemented by two threshold Cherenkov counters {\small \v{C}$_{1}$, \v{C}$_{2}$} for kaon identification 
and by beam  scintillation counters 
{\small S$_{1}$}, {\small  S$_{2}$}, {\small S$_{3}$}, {\small S$_{4}$}. 
The 11~m long {\small Decay Volume (DV)} filled with helium contains 11 rings of guard system ({\small GS}).
To reinforce {\small GS}, a gamma detector ({\small BGD}), made of lead glass blocks  
located behind the {\small DV} is used as a veto at large angles, while low angle particles pass through a central opening. 
The wide aperture  spectrometric magnet, 
{\small SM$_{(SP40A)}$}, with a field integral of $\sim$~1~Tm serves as a spectrometer for the charged 
decay products together with corresponding tracking chambers: 2~mm PC's ({\small PC$_{1,...,8}$}), 
9~mm diameter straw tubes {\small ST$_{(1,2,3)}$} and 30~mm diameter drift tubes {\small DT$_{1,2}$}.
The matrix hodoscope {\small HODO$_{(matrix)}$} is composed of 252 scintillator tiles with WLS+SiPM readout.
It is used in the trigger, improves time resolution and links $x$--$y$ projections of a track.
Two scintillator counters {\small S$_{bk1}$, S$_{bk2}$} serve to suppress undecayed beam particles.
At the end of the OKA setup there are two calorimeters: electromagnetic ({\small GAMS-2000}) made of 
lead glass blocks and a hadron one ({\small HCAL$_{(GDA)}$})- 100 iron-scintillator sandwiches. Finally, four partially overlapping muon counters are located  downstream the {\small HCAL}.
The data acquisition system of the OKA setup \cite{aFilinOkaDAQ} operates at $\sim$~25~kHz event rate
with the mean event size of $\sim$~4~kByte. The detailes of the setup and the beam can be found elswhere \cite{Kmunuh}.
%%----fixed-----

\section{The analysis procedure}
The study of the decay  $K^{+}\to\pi^{+}\pi^{-}\pi^{+}\gamma$ is performed with the data set accumulated in 2012 and 2013 
runs with  17.7 GeV/c beam momentum. Two  triggers were used.
The first one selects beam kaons decays inside the OKA setup:
${\tt Tr_{Kdecay}=S_{1}{\cdot}S_{2}{\cdot}S_{3}{\cdot}S_{4}{\cdot}\check{C}_{1}{\cdot}\overline{\check{C}}_{2}{\cdot}\overline{S}_{bk}}$ and, in addition, requires
an energy deposition in GAMS-2000 e.m. calorimeter higher than MIP: ${\tt Tr_{Gams}=Tr_{Kdecay} \cdot (E_{Gams}>MIP)}$.
The second one, ${\tt Tr_{HODO}=Tr_{Kdecay}{\cdot}{(2 \leq Mult \leq 4})}$, includes additionally
a requirement on multiplicity in the Matrix hodoscope.
The beam intensity ({\tt S$_{1}\cdot$S$_{2}\cdot$S$_{3}\cdot$S$_{4}$}) was $\sim 2\cdot 10^{6}$ per spill, 
the fraction of kaons in the beam was $\sim$ 12.5\%, i.e.~the kaon intensity was  $\sim$250k/spill. 
The total number of kaons entering the DV corresponds to $\sim 3.4\times 10^{10}$.  
For the estimation of the background contribution to the selected data set, a sample of the Monte Carlo events with  six main decay channels of charged kaon ( $\pi^+\pi^0$, $\pi^+\pi^0\pi^0$, $\pi^+\pi^-\pi^+$, $\mu^+\nu$, $\pi^0\mu^+\nu$, $\pi^0e^+\nu$) mixed accordingly to the branching fractions, with the total statistics about  equal to that of  the recorded data sample is used.  The Monte Carlo events are passed through  full OKA simulation and reconstruction procedures. Monte Carlo sample  for the signal events is  produced in the same way as for the background. As the input, the weight,  proportional to the square of the absolute value of the matrix element for the decay $K^{+}\to\pi^{+}\pi^{-}\pi^{+}\gamma$, given in \cite{Ambr1} is used. 
 
\subsection{Event selection}
To select $K^{+}\to\pi^{+}\pi^{-}\pi^{+}\gamma$ decay channel in off-line analysis a set of requirements is applied:\\ 
-- the momentum of the beam track is required to be measured;\\
-- the number of the secondary charged tracks is required to be equal to three and the net charge of them is equal to +1;\\
-- the decay vertex should have good $\chi^2$ and should be with a margin incide the DV ;\\
-- the charged tracks are not identified as electrons in the GAMS-2000 electromagnetic calorimeter;\\
-- the  missing mass squared to each positive pion $M^{2}_{miss}(\pi^+)=(P_{K^{+}}-P_{\pi^{+}})^2$ should be greater than 0.07 GeV$^2$;\\
-- the event should contain one and only one  photon with the energy $E_\gamma >0.5$ GeV
-- the invariant mass of the photon with each pion ($M(\pi\gamma)$)should be  greater than 0.17 GeV ;\\
-- the square of the transverse momentum of the  $\pi^+\pi^-\pi^+\gamma$ system  is less than 0.001 GeV$^2$;\\ 
-- the ratio of the momentum of the  $\pi^+\pi^-\pi^+\gamma$ system  to the momentum of the beam track should be within the range 0.95-1.05.\\

The cut on the square of the missing mass to each positive pion $M^{2}_{miss}(\pi^+)$ is used for the suppression of the background from 
the decay $K^{+}\to\pi^{+}\pi^{0}$ with $\pi^{0}\to e^{+}e^{-}\gamma$.%% The distribution on $M^{2}_{miss}(\pi^+)$ is shown in Fig~\ref{mm2_pi} 
%%for the data, main sources of the background and the signal. The distribution for the background reproduces well the peak at the square 
%%of $\pi^0$ mass, visible in the data. 
The main source of the background for the decay $K^{+}\to\pi^{+}\pi^{-}\pi^{+}\gamma$ is the decay of kaon to three charged pions 
when the pions produce hadron showers in the  electromagnetic calorimeter and, because of fluctuations, a part of a shower is not 
associated with a charged track by the reconstruction program. To suppress such kind of background a cut on $M(\pi\gamma)$ is done. 
The reconstructed momentum of the $\pi^{+}\pi^{-}\pi^{+}\gamma$ system should be equal to the momentum of the beam particle. 
This motivates the last two cuts listed above. %%The distributions of the square of the transverse momentum of the $\pi^{+}\pi^{-}\pi^{+}\gamma$ 
%%system and the ratio of absolute value of its momentum to the momentum of the beam particle is shown in Fig.~\ref{p2t} and Fig.~\ref{xb}. 

\begin{figure}[!ht]
\begin{center}
\includegraphics[width=0.55\linewidth]{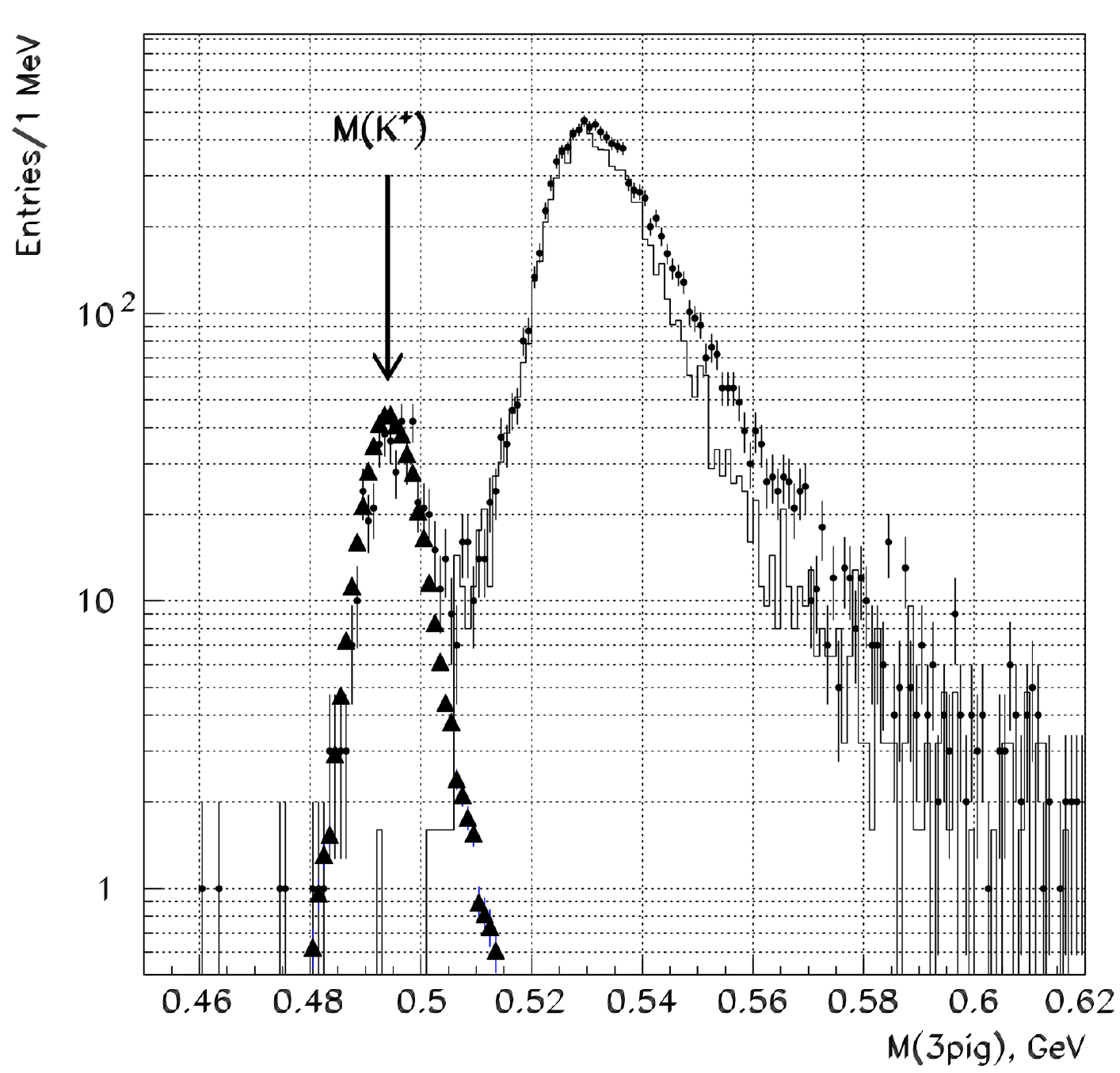}
\caption{
  The invariant mass distribution of the $\pi^{+}\pi^{-}\pi^{+}\gamma$ system for the data (black circles), 
  main sources of background (histogram) and signal decays of the kaons (black triangles).
}
\label{m3pig}
\end{center}
\end{figure}

The invariant mass distribution of the $\pi^{+}\pi^{-}\pi^{+}\gamma$ system after application of  all  the cuts listed above is shown in Fig.~\ref{m3pig}. We see clear separation of the signal, peaking around the nominal value of the kaon PDG~\cite{PDG} mass and the background concentrating at higher masses. The number of events in the signal region is about 450.
As almost all the photons in the selected $K^{+}\to\pi^{+}\pi^{-}\pi^{+}\gamma$ events have the energy in the kaon rest frame  greater than 0.03 GeV, in the following analysis we apply an additional explicit cut $E_{\gamma}^{*} > 0.03$ GeV.

\subsection{Measurement of the branching ratio  of the $K^{+}\to\pi^{+}\pi^{-}\pi^{+}\gamma$ decay}
\begin{figure}[!ht]
\begin{minipage}{0.47\linewidth}
\centering
\includegraphics[width=1.01\linewidth]{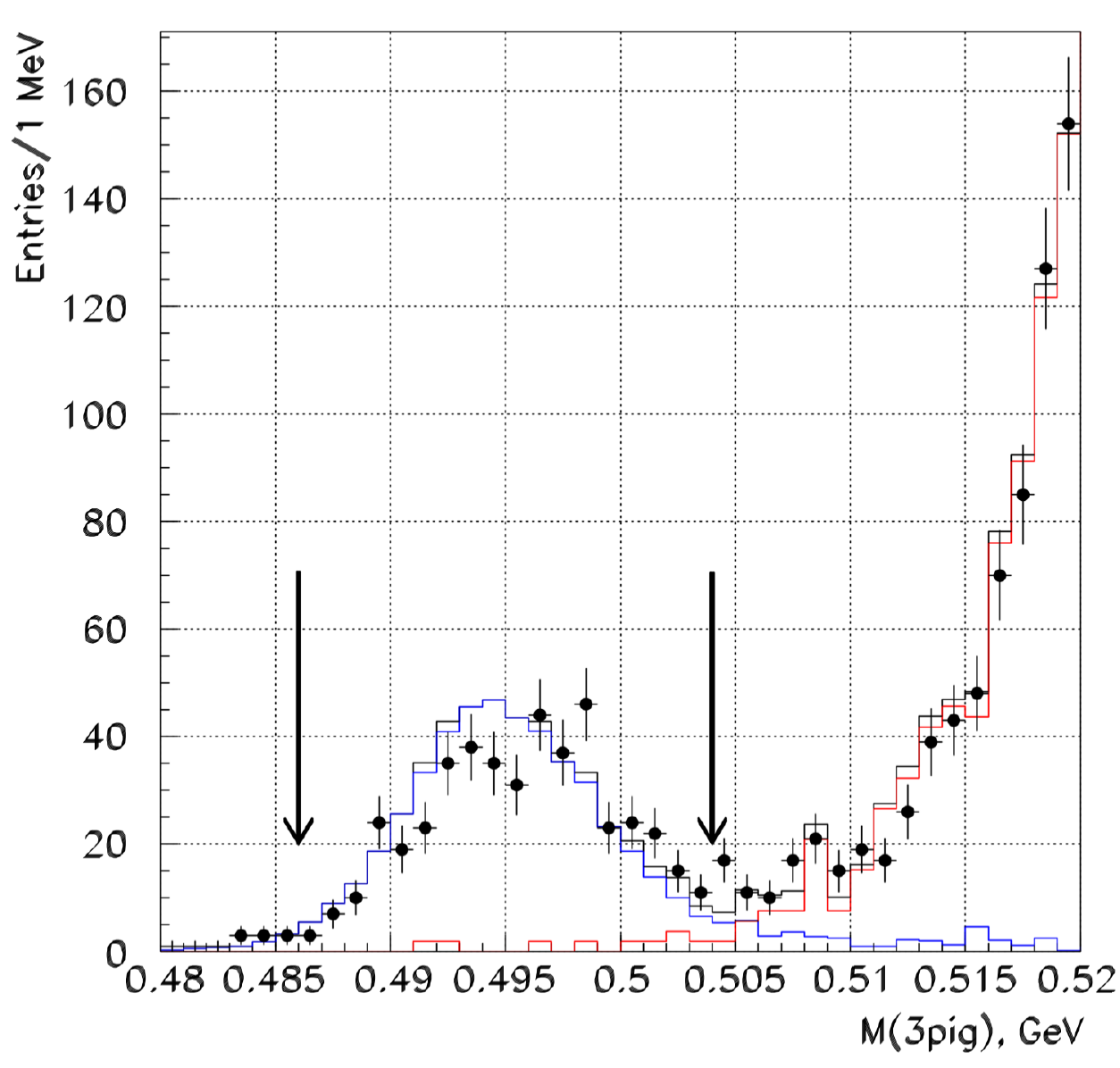}
\end{minipage}
\begin{minipage}{0.05\linewidth}
~\\
\end{minipage}
\begin{minipage}{0.47\linewidth}
\centering
\includegraphics[width=1.00\linewidth]{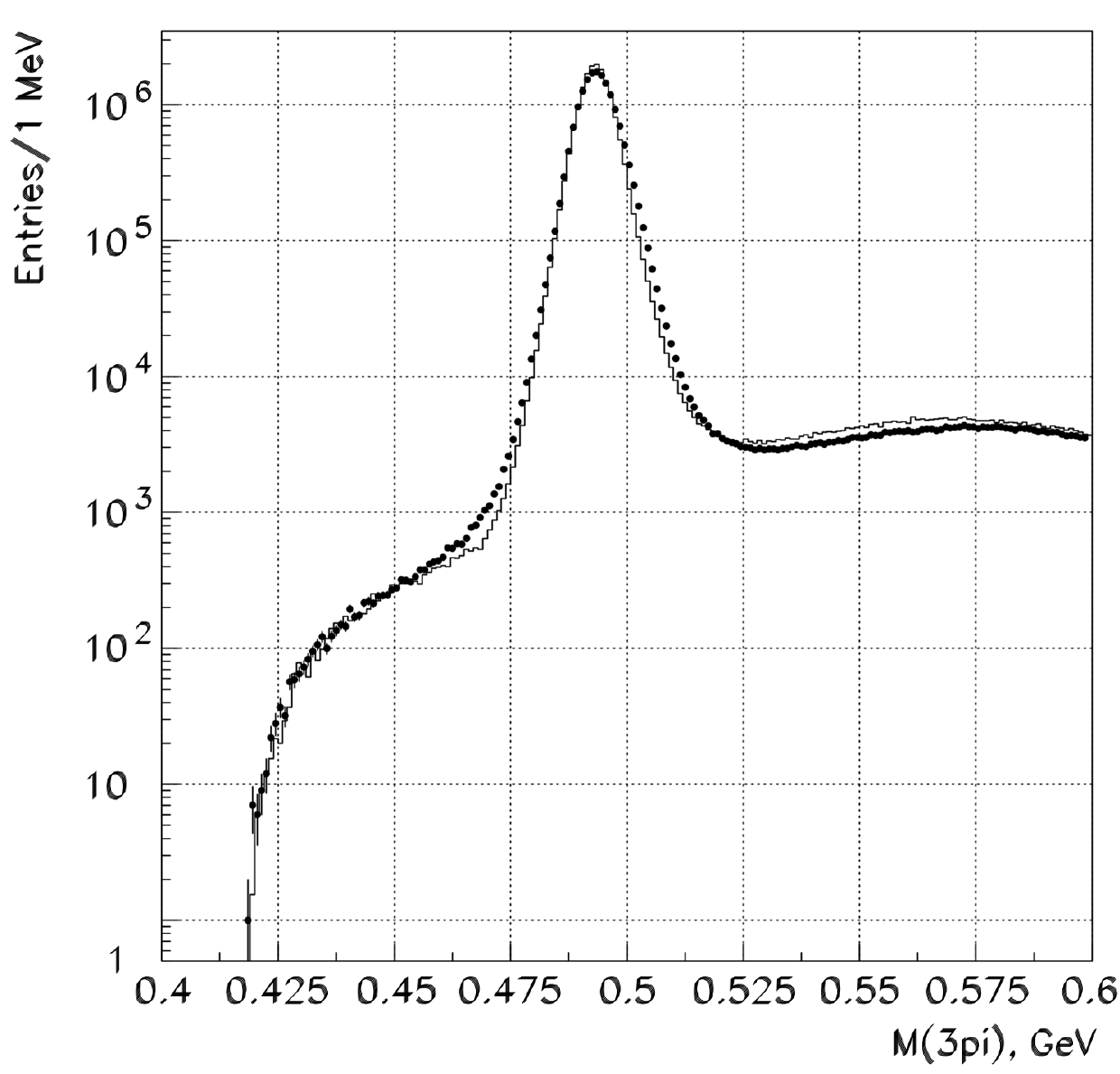}
\end{minipage}\\
\begin{minipage}{0.48\linewidth}
\vspace{4mm}
\caption{
  The invariant mass distribution of the $\pi^{+}\pi^{-}\pi^{+}\gamma$ system for
  the data (black circles), the  signal decays (blue histogram), the main sources 
  of the background (red histogram) and the sum of the contributions of the signal decays 
  and the  background (black histogram). The arrows show the signal region used in the analysis.
  \label{m3pig2}
}
\end{minipage}
\begin{minipage}{0.05\linewidth}
~\\
\end{minipage}
\begin{minipage}{0.46\linewidth}
\vspace{4mm}
\caption{
  The invariant mass distribution of  three charged pions for 
  the data (black circles) and  for the Monte Carlo  for the six main  
  channels of the charged kaon decay (histogram).
  \vspace{12.0mm}
  \label{m3pi1}
}
\end{minipage}
\end{figure}

The invariant mass distribution of the $\pi^{+}\pi^{-}\pi^{+}\gamma$ system in the  signal region is shown in Fig.~\ref{m3pig2}. The Monte Carlo distributions for the main sources of the background and the signal events are shown together with the data. We define the mass range of 0.486-0.504 GeV as a signal region. The number of the decays of $K^{+}\to\pi^{+}\pi^{-}\pi^{+}\gamma$ is determined as the difference of the number of the data events in the signal region and the expected number of events from the background. The expected background contribution to the signal region is determined using Monte Carlo events for six main channels of the kaon decays described in the above section.  
To determine the branching fraction of the decay $K^{+}\to\pi^{+}\pi^{-}\pi^{+}\gamma$ we use 
the decay $K^{+}\to\pi^{+}\pi^{-}\pi^{+}$ as  the normalization channel. We expect the cancellation 
of many ambiguities in the ratio of the branching fractions Br($K^{+}\to\pi^{+}\pi^{-}\pi^{+}\gamma$)/Br($K^{+}\to\pi^{+}\pi^{-}\pi^{+}$). 
The invariant mass spectrum of the system of three charged pions for the same selection criteria as for the charged part 
of the $\pi^{+}\pi^{-}\pi^{+}\gamma$ decay  is shown in  Fig~\ref{m3pi1}. 

\vspace{7.0mm}
\vspace{7.0mm}

\begin{figure}[!ht]
\begin{center}
\includegraphics[width=0.95\linewidth]{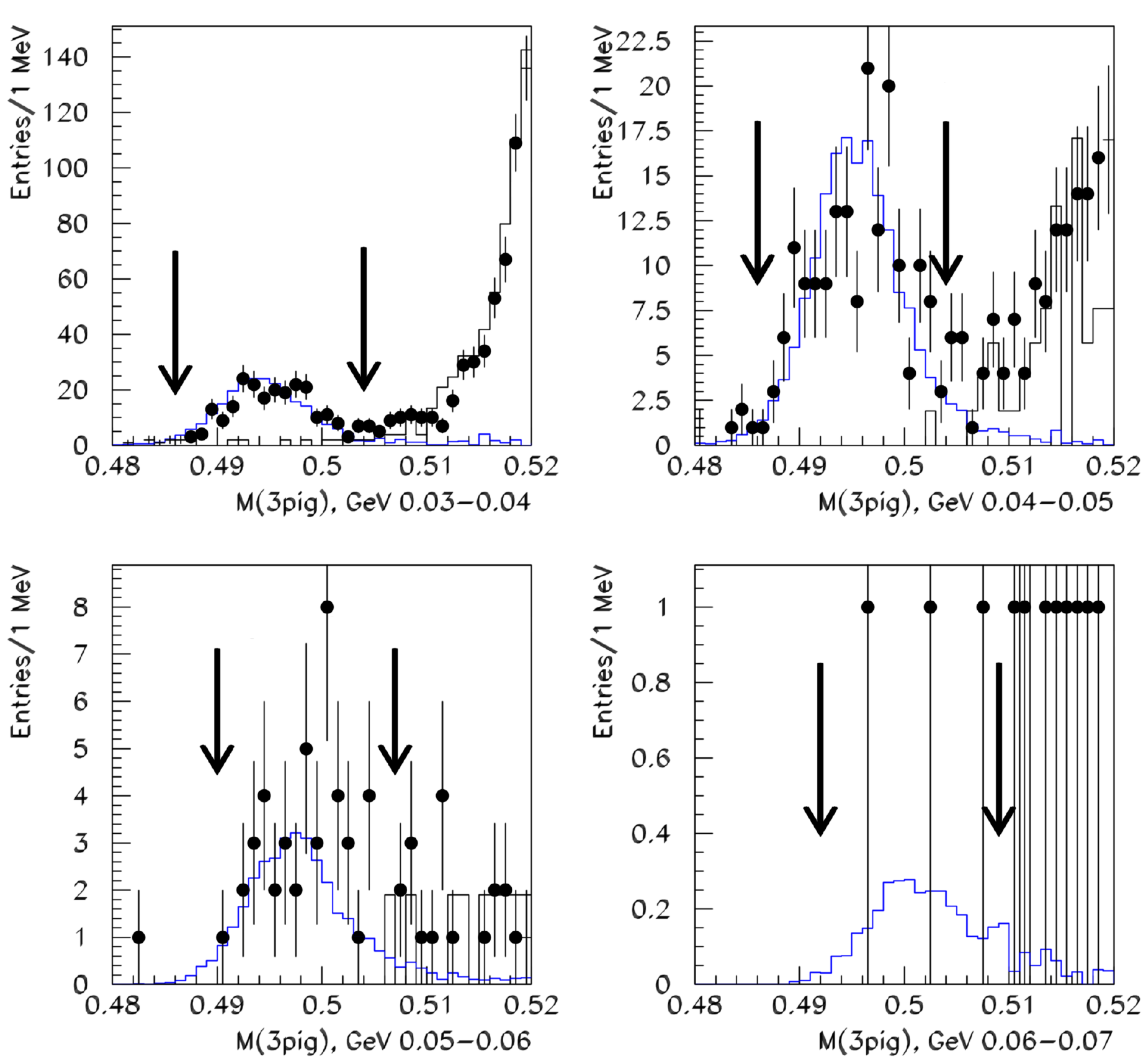}
\caption{
  Invariant mass distributions of the $\pi^{+}\pi^{-}\pi^{+}\gamma$ system for
  the data (black circles), signal decays (blue histogram) and main sources of the background (black histogram) for the different energy ranges 
  of the photon in the decay $K^{+}\to\pi^{+}\pi^{-}\pi^{+}\gamma$. The top left distribution is for the photon energy range 0.03-0.04 GeV, 
  the top right one is for the range 0.04-0.05 GeV, the bottom left one is for the 0.05-0.06 GeV and the last one is for the 0.06-0.07 GeV. 
  The arrows show the signal regions used in the analysis.
}
\label{m3pig3}
\end{center}
\end{figure}
\begin{figure}[!ht]
\begin{minipage}{0.47\linewidth}
\centering
\includegraphics[width=1.01\linewidth]{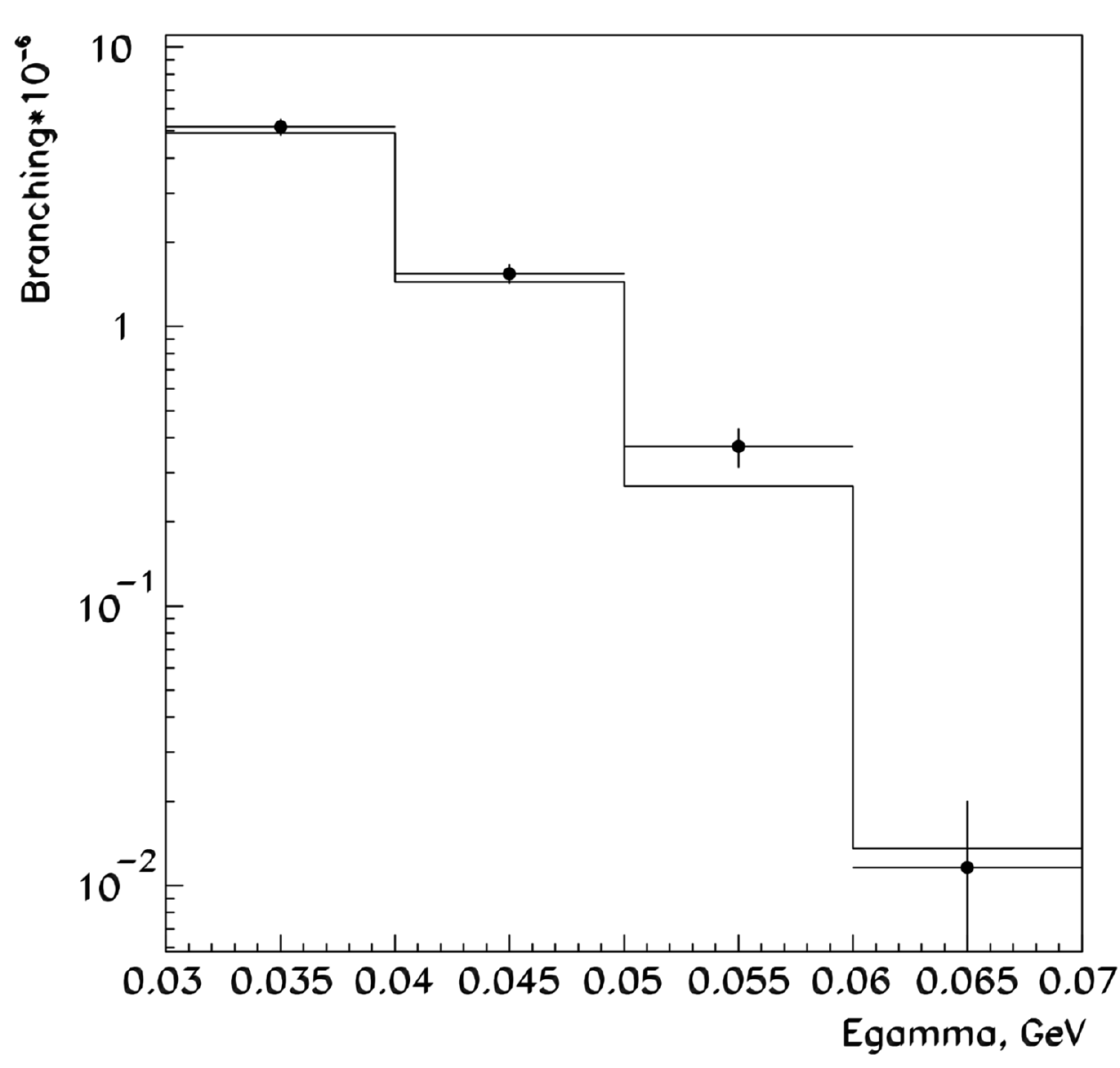}
\end{minipage}
\begin{minipage}{0.05\linewidth}
~\\
\end{minipage}
\begin{minipage}{0.47\linewidth}
\centering
\includegraphics[width=1.01\linewidth]{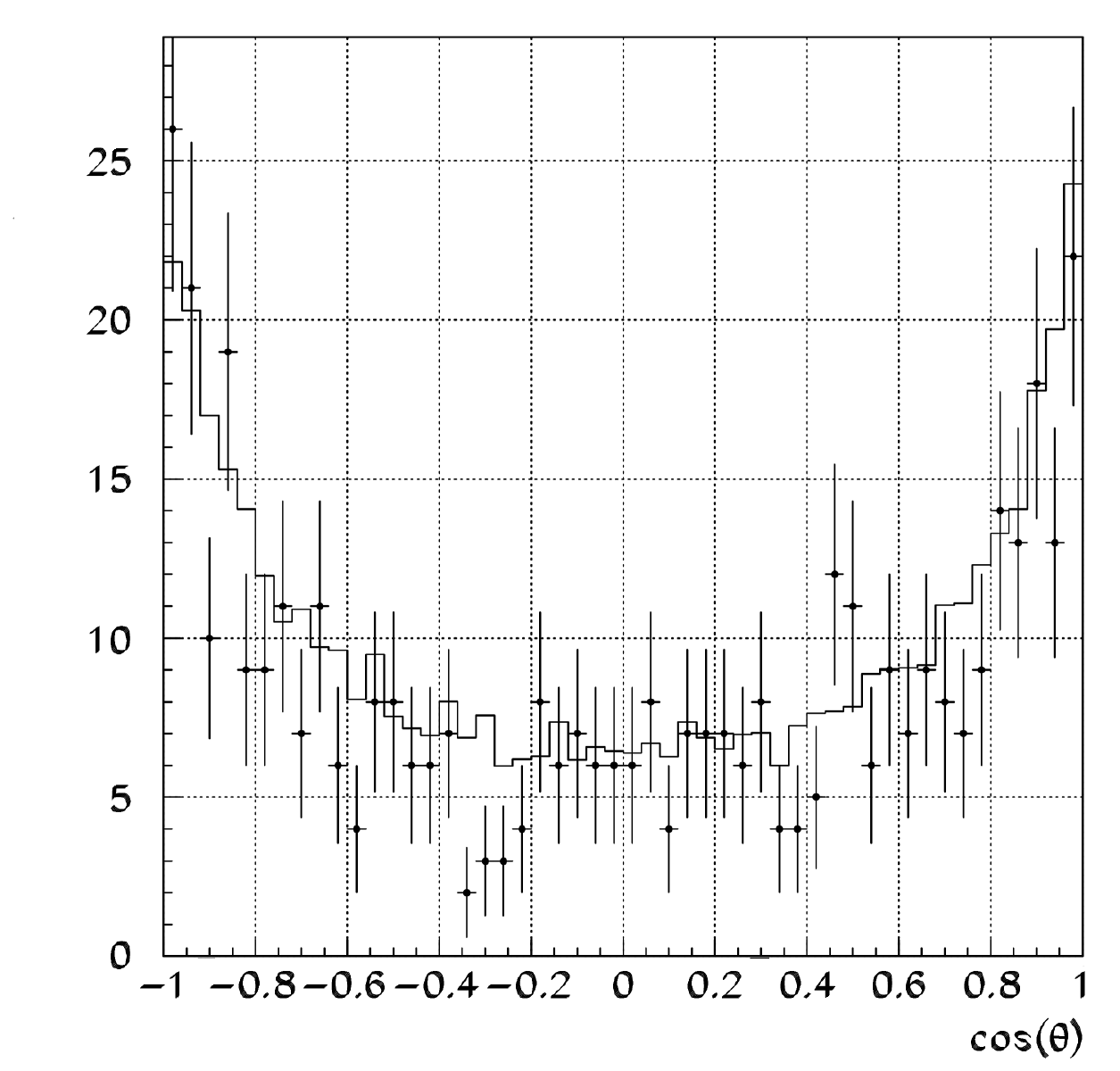}
\end{minipage}\\
\begin{minipage}{0.47\linewidth}
\caption{
  Measured branching fractions for the decay  $K^{+}\to\pi^{+}\pi^{-}\pi^{+}\gamma$ in the photon 
  energy ranges in the kaon rest frame  for the data (black circles)
  and the CHPT prediction from \cite{Ambr2} (histogram).
  \label{br_egam}
}
\end{minipage}
\begin{minipage}{0.05\linewidth}
~\\
\end{minipage}
\begin{minipage}{0.47\linewidth}
\vspace{10.4mm}
\caption{
  The distribution (black circles) of the cosine of the angle of the photon direction with respect 
  to the  hadronic system decay plane(see the text). The histogram shows the Monte Carlo prediction 
  for the decay $K^{+}\to\pi^{+}\pi^{-}\pi^{+}\gamma$ with the matrix element from \cite{Ambr2}.
  \label{ngpfps}
}
\end{minipage}
\end{figure}
\vspace{10mm}
 The branching fraction of the decay $K^{+}\to\pi^{+}\pi^{-}\pi^{+}\gamma$ is determined in the following way:
$$
Br(3\pi\gamma)=Br(3\pi)_{PDG}\times \epsilon(3\pi) \times N(3\pi\gamma)_{D}/\epsilon(3\pi\gamma)/N(3\pi)_{D},
$$
where $Br(3\pi)_{PDG}$ is the branching fraction of the decay $K^{+}\to\pi^{+}\pi^{-}\pi^{+}$ from PDG~\cite{PDG}; $\epsilon(3\pi)$ and $\epsilon(3\pi\gamma)$ are the  reconstruction efficiencies for the decays $K^{+}\to\pi^{+}\pi^{-}\pi^{+}$ and $K^{+}\to\pi^{+}\pi^{-}\pi^{+}\gamma$, determined from the Monte Carlo; $N(3\pi\gamma)_{D}$ and $N(3\pi)_{D}$ are the numbers of the decays of $K^{+}\to\pi^{+}\pi^{-}\pi^{+}\gamma$ and $K^{+}\to\pi^{+}\pi^{-}\pi^{+}$ in the data. Efficiencies $\epsilon(3\pi)$ and $\epsilon(3\pi\gamma)$, determined from the Monte Carlo, are $0.12 \pm 0.002 $ and $0.024 \pm 0.001$, respectively. The obtained result is
$$
Br(3\pi\gamma)=(0.71\pm 0.04(stat.))\times 10^{-5}.
$$

The main source of the systematic error is the uncertainty in the estimate of the background contribution to the signal region (see Fig.~\ref{m3pig},\ref{m3pig2}). For the estimation of this uncertainty we varied the normalization of the background distribution.The variations are  quadratically summed with that due to the changes of the cuts listed in the selection criteria. The estimated value of the systematic uncertainty for the branching fraction is $0.03\times 10^{-5}$. The final result is
$$
Br(3\pi\gamma)=(0.71\pm 0.04(stat.)\pm 0.03(syst.))\times 10^{-5}, E^*_{\gamma} > 0.03 GeV.
$$

The theory prediction in the framework of the CHPT is $Br_{th}=(0.665\pm0.005)\times 10^{-5}$~\cite{Ambr2}.

\subsection{Measurement of the photon energy spectrum}  

To perform a measurement of the photon energy spectrum, we split the data sample into four parts with photon energies in the rest frame of the kaon lying in the ranges 0.03-0.04, 0.04-0.05, 0.05-0.06 and 0.06-0.07 GeV. After that we apply the procedure of the previous section to each of the four data samples. The $\pi^{+}\pi^{-}\pi^{+}\gamma$ invariant mass spectra for the listed above ranges of the photon energy are shown in Fig~\ref{m3pig3} together with expected signal and background contributions.

The obtained values of the branching fractions for the given energy ranges  are listed in the table 1 and shown in the Fig.~\ref{br_egam}. For comparison, we also give the {\cal O}$(p^{4})$ CHPT predictions from~\cite{Ambr2}. 

\begin{table*}
\caption{Values of the branching fractions for the decay $K^{+}\to\pi^{+}\pi^{-}\pi^{+}\gamma$ in the  intervals of the photon energy  in the kaon rest frame. The uncertainties of the data given in the table are the statistical ones.
\label{tab:br}}
\begin{center}\begin{tabular}{c|c|c} \hline \hline
{\bf\small Energy interval (GeV)}&{\bf\small Branching fraction(data)}&{\bf\small Branching fraction(CHPT~\cite{Ambr2})}\\ \hline
0.03-0.04  & $(5.17 \pm 0.34)\cdot 10^{-6}$ & $(4.93 \pm 0.05)\cdot 10^{-6}$ \\
0.04-0.05  & $(1.55 \pm 0.12)\cdot 10^{-6}$ & $(1.44 \pm 0.01)\cdot 10^{-6}$ \\
0.05-0.06  & $(0.35 \pm 0.05)\cdot 10^{-6}$ & $(0.269 \pm 0.003)\cdot 10^{-6}$ \\
0.06-0.07  & $(0.11 \pm 0.06)\cdot 10^{-7}$ & $(0.136 \pm 0.002)\cdot 10^{-7}$ \\ \hline \hline
\end{tabular}\end{center}
\end{table*}

\section{Search for  a photon up-down asymmetry }

In a decay in the beauty sector $B^{+}\to K^{+}\pi^{+}\pi^{-}\gamma$ the LHCb experiment has found significant up-down asymmetry of the photon  with respect to the  hadronic system decay plane~\cite{Pol1}.  This observable was proposed in~\cite{Pol2}, it is both P and T-odd. We perform  an analogous study for the radiative kaon decay to three charged pions and photon to search for a New Physics effects. In Fig~\ref{ngpfps} we show the distribution of the cosine of the angle of the photon direction with respect to the  pion system decay plane:
$$
cos(\theta) = n_\gamma \cdot [p_{f}(\pi) \times p_{s}(\pi)]/|[p_{f}(\pi) \times p_{s}(\pi)]|, 
$$
where $n_\gamma$ is the unit vector of photon direction in the 3-pion rest frame, $[p_{f}(\pi) \times p_{s}(\pi)]$ is the vector product of the momenta of the fastest and  slowest pions in the same frame. For comparison, we show the same distribution for the Monte Carlo signal events with the matrix element from~\cite{Ambr2}. The observed asymmetry in the data
$$ 
A = (N(cos\theta > 0) - N(cos\theta < 0))/N_{total} = 0.03 \pm 0.05 \pm 0.03
$$
is consistent with zero within statistical error.

\newpage
\section*{Conclusions}
The decay $K^{+}\to\pi^{+}\pi^{-}\pi^{+}\gamma$ is studied on statistics of ~450 events. 
The measured branching fraction is $Br(3\pi\gamma) = (0.71 \pm 0.05)\times 10^{-5}$ for $E_\gamma^* > 0.03$ GeV. The photon energy spectrum  for this decay is also determined. The measured branching fraction and the energy spectrum agree well within the errors with the calculations in the framework of the chiral perturbation theory. The measured up-down asymmetry of the photon with respect to the decay plane of the  hadronic system is $0.03\pm 0.06$. No sign of P and T-odd effects is observed.

\section*{Acknowledgements}
We express our gratitude to our colleagues in the accelerator department for the good performance of the U-70 during data taking; 
to colleagues from the beam department for the stable operation of the 21K beam line, including RF-deflectors, and to colleagues 
from the engineering physics department for the operation of the cryogenic system of the RF-deflectors.

The work is supported  by the Russian Fund for Basic Research, grant N18-02-00179A.
%\clearpage

\end{document}